\documentclass{ws-procs975x65}

\begin{document}

\title{\uppercase{Effects of Compton Cooling on outflows in a Two 
Component Accretion Flow around a Black Hole: Results of a Coupled 
Monte Carlo-TVD Simulation}}

\author{SUDIP K. GARAIN $^{1,*}$, HIMADRI GHOSH $^2$ and SANDIP K. CHAKRABARTI $^{1,2}$}

\address{$^1$ S.N. Bose National Centre for Basic Sciences,
JD Block, Salt Lake, Sector III, Kolkata, 700098, India\\
$^2$ Indian Centre for Space Physics, Chalantika 43, Garia Station Rd.,
Garia, Kolkata, 700084,India\\	
$^*$E-mail: sudip@bose.res.in}

\begin{abstract}
The effect of cooling on the outflow rate from an accretion disk around
a black hole is investigated using a coupled Monte Carlo Total Variation 
Diminishing code. A correlation between the spectral states and the 
outflow rates is found as a consequence.
\end{abstract}


\bodymatter

\section{Introduction}
Outflows are common in many astrophysical objects which contain black holes. 
It is generally believed that the outflows originate from the inflowing gas itself. In a numerical
simulation using total variation diminishing (TVD) code, Giri et al. (2010) 
showed that the outflow rates from an inviscid accretion flow strongly 
depend on the outward centrifugal force and upto 60\%
matter can be driven out of the disk. Here, we discuss how the Compton 
cooling affects the outflow from the accretion disk.
\section{Simulation Set Up and Procedure}
\begin{figure}[h!]
\vskip -1.5cm
\centering{
\includegraphics[height=7truecm,width=7truecm,angle=0]{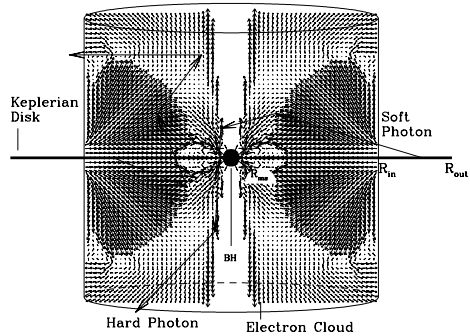}}
\vskip -1.5cm
\caption{\footnotesize{Schematic diagram of the geometry of our 
radiative hydrodynamics simulations. Zigzag trajectories are the 
typical paths followed by the photons. The velocity
vectors of the infalling matter inside the cloud are shown. The velocity vectors
are plotted for $\lambda=1.73$.}}
\end{figure}
\vskip -0.2cm
In Figure 1, the schematic diagram of our simulation set up is presented.
The outer boundary of the sub-Keplerian matter is at $R_{in} = 100 r_g$ 
whereas that of the Keplerian disk at the equatorial plane is located 
at $R_{out} = 200 r_g$. At the center, a non-rotating black hole of 
mass 10$M_{\odot}$ is located. We have calculated the flow dynamics 
using the TVD code (Ryu et al. 1997; Giri et al. 2010).
For a particular simulation, we use the Keplerian disk rate ($\dot{m}_d$) 
and the sub-Keplerian halo rate ($\dot{m}_h$) as parameters. The specific 
energy ($\epsilon$) and the specific angular momentum ($\lambda$) 
determines the hydrodynamics (shock location, number
density and velocity variations etc.) and the thermal properties 
of the sub-Keplerian matter.

The radiative properties of the accretion disk is studied using a Monte 
Carlo code (Poznyakov et al. 1983; Ghosh et al. 2009).
Each photon from the Keplerian disk is tracked
till either it leaves the accretion disk or is absorbed by the black 
hole. If a scattering between an electron and a photon occurs on the way, 
the amount of energy exchanged is computed and suitably adjusted against 
the total energy of the electron and the photon. \\
\indent The hydrodynamic code and the radiative transfer code is 
coupled together and are run back to back. In this
way, we get an opportunity to see how the dynamics of the flow changes 
because of the Compton cooling and how the
radiation spectrum from the accretion disk vary with time.
\section{Results and Discussions}
\begin{center}
{\footnotesize{
\begin {tabular}[h]{cccc}
\multicolumn{4}{c}{Table 1: Parameters used for the simulations.}\\
\hline Case & $\epsilon, \lambda$ &  $\dot{m}_h$ & $\dot{m}_d$ \\
\hline
Ia & 0.0021, 1.76 & 1.0 & No Disk \\
Ib & 0.0021, 1.76 & 1.0 & 0.5     \\
Ic & 0.0021, 1.76 & 1.0 & 1.0     \\
Id & 0.0021, 1.76 & 1.0 & 2.0     \\
\hline
IIa & 0.0021, 1.73 & 1.0 & No Disk  \\
IIb & 0.0021, 1.73 & 1.0 & 0.5      \\
IIc & 0.0021, 1.73 & 1.0 & 1.0      \\
IId & 0.0021, 1.73 & 1.0 & 2.0      \\
\hline
\end{tabular}
}}
\end{center}
In Table 1, we list various Cases with all the simulation parameters used.
In Cases Ia and IIa, there is no Keplerian disk. These are non-radiative 
hydro-dynamical simulations and no Compton cooing is included.

We find that the average temperature in the post-shock region is reduced 
rapidly as the disk rate $\dot{m}_d$ is increased keeping the halo rate 
$\dot{m}_h$ fixed for both the angular momenta cases. 
As a result, the average shock location decreases since cooling reduces 
the thermal pressure in this region.
\begin{figure}
\begin{center}
\vskip -0.5cm
\includegraphics[width=4.0cm,height=4.0cm]{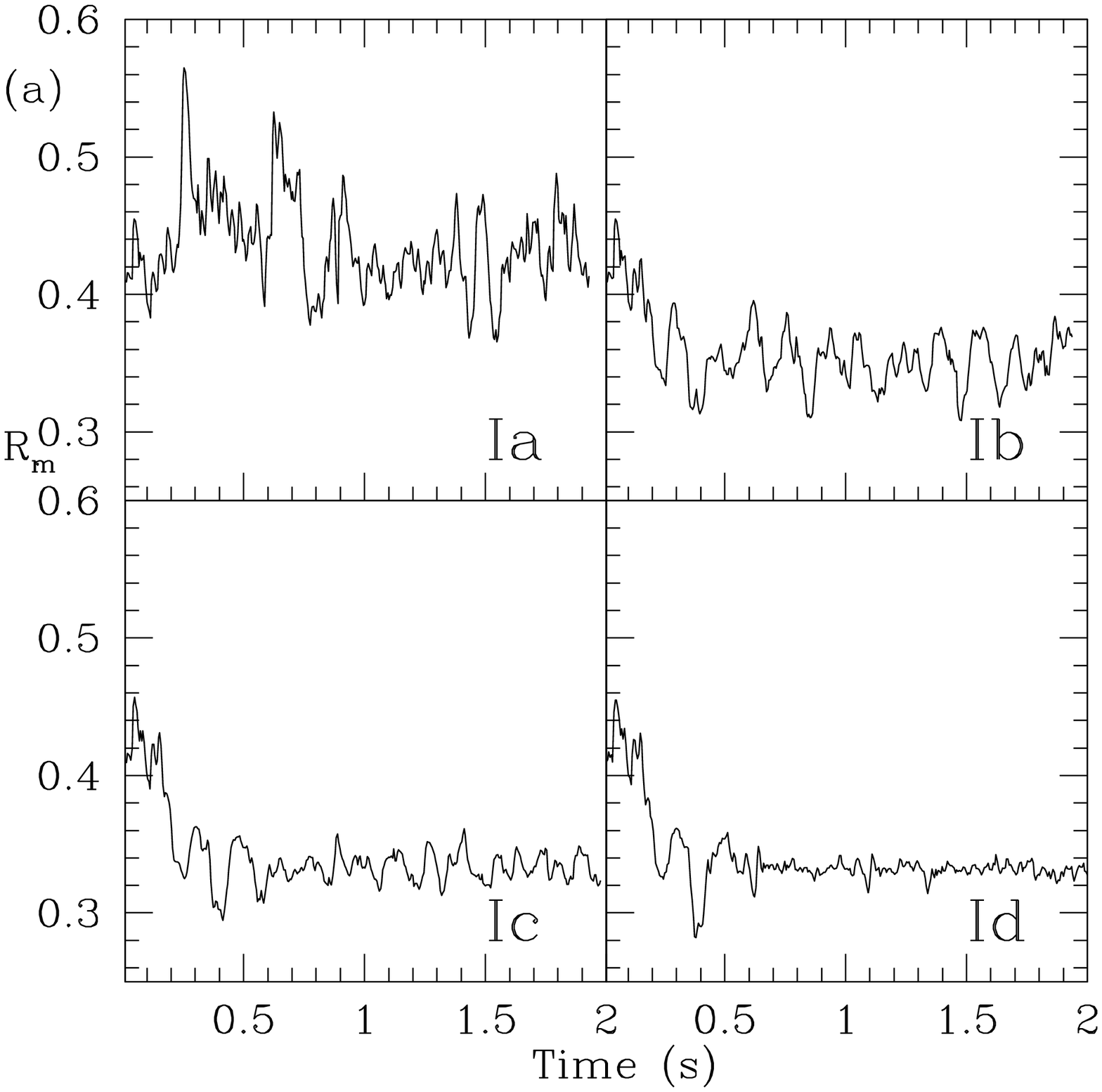}
\includegraphics[width=4.0cm,height=4.0cm]{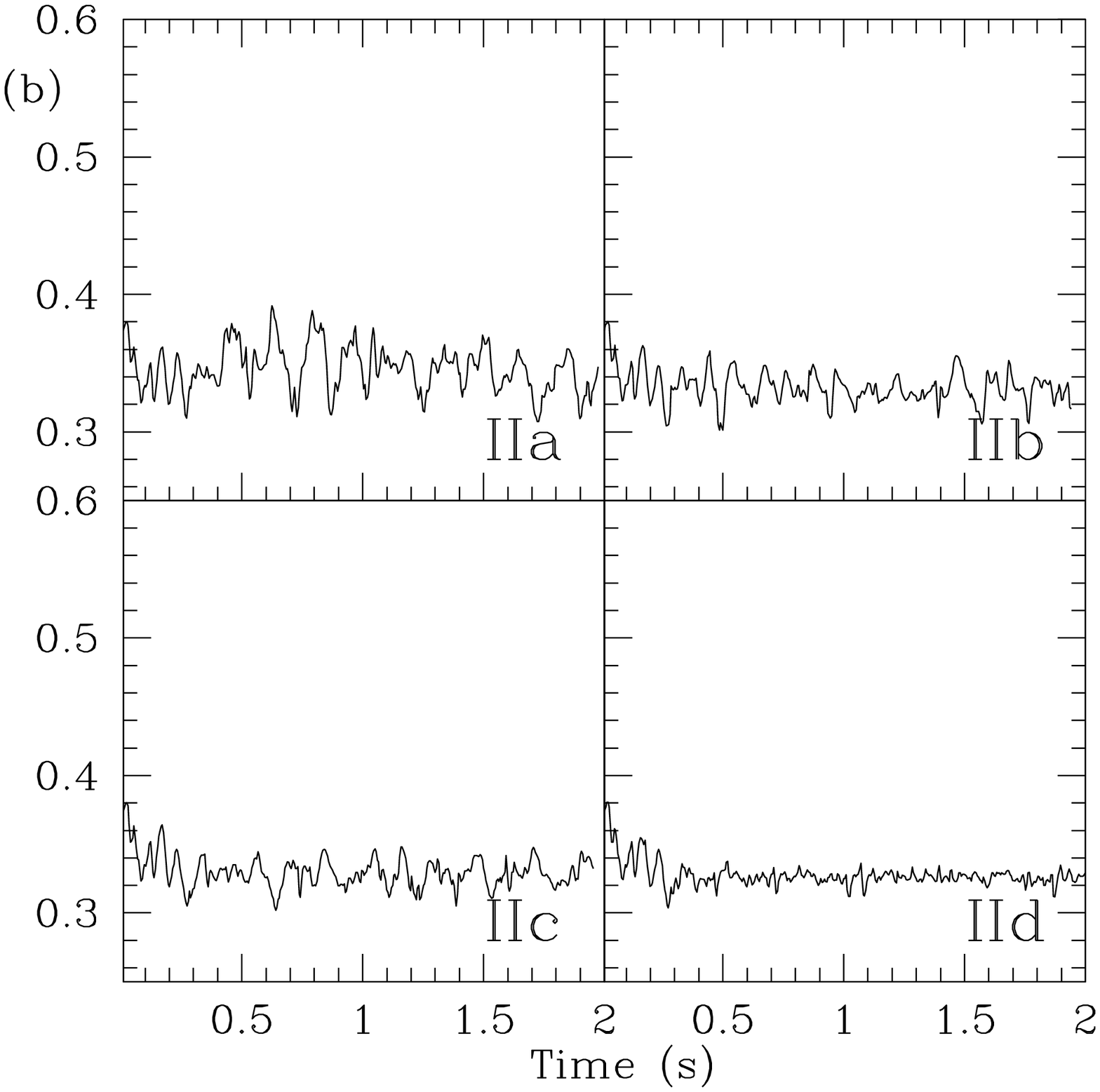}
\vskip -0.2cm
\caption{\footnotesize{Variations of $R_{\dot{m}} (=\frac{\dot{M}_{out}}
{\dot{M}_{in}})$ with time for different $\dot{m_d}$ is shown here. 
(a) $\lambda = 1.76$ and (b) $\lambda = 1.73$.
The Cases are marked in each panel. The outflow rate is the lowest for the
highest Keplerian disk accretion rate (Cases are Id and IId).}}
\end{center}
\end{figure}
We now study how the Comptonization affects the outflow rates.
We define the outflow rate $\dot{M}_{out}$ to be the rate at 
which outward pointing flow leaves the computational grid. 
In Figure 2, we plot the results of the time variation of  
the ratio $R_{\dot{m}}$($\frac{\dot{M}_{out}}{\dot{M}_{in}}$), 
${\dot{M}_{in}}$ being the constant injection rate on the right 
boundary. It can be clearly observed that with the 
increase in cooling the ratio is dramatically reduced and indeed 
becomes almost saturated as soon as some cooling is introduced.\\
\begin{figure}
\begin{center}
\vskip -0.5cm
\includegraphics[width=7cm,height=3.5cm]{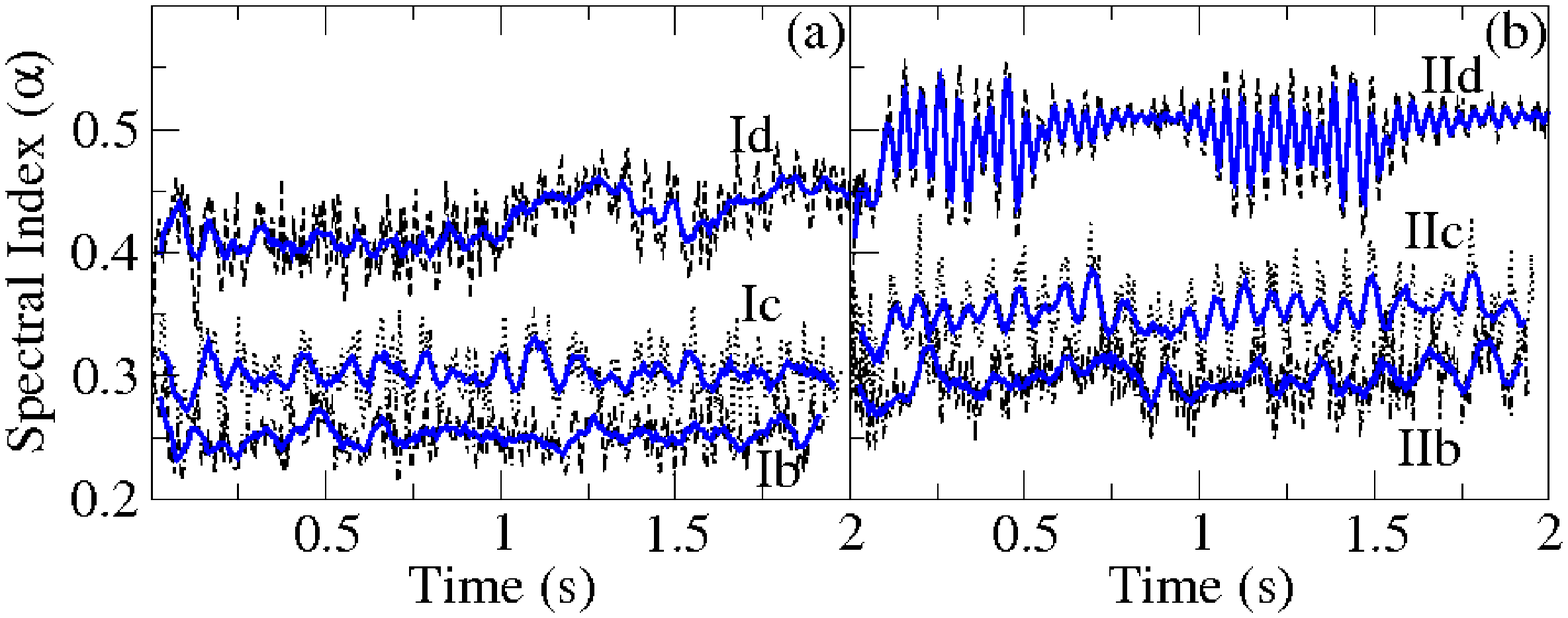}
\caption{\footnotesize{Time variation of the spectral slope ($\alpha$, 
$I(E) \propto E^{-\alpha}$) for different disk rates and their running 
averages (solid line) are shown. Different Cases are marked.}}
\vskip -0.5cm
\end{center}
\end{figure}
\indent The time variation of spectral properties of the disk is also 
studied. Whether the spectrum will be soft or hard is determined by the 
relative availability of the hot electrons and the soft
photons in the disk and the jet. As the increase in $\dot{m}_d$ cools 
the post shock region and shrinks its size, the spectrum also becomes 
softer. If we define the energy spectral index  $\alpha$ to be
$I(E) \propto E^{-\alpha}$ in the region $1-10$ keV, we note that 
$\alpha$ increases with the increase in $\dot{m}_d$. In Figure 3, 
we present the time variation of the spectral index for the different
Cases. We draw the running mean through these variations.  We clearly 
see that the spectral index goes up with the increase in the disk 
accretion rate. Thus, on an average, the spectrum softens.
We also find that the spectrum oscillates quasi-periodically and the 
frequency is higher for higher cooling rate. This agrees with the 
general observations that the quasi-periodic oscillation (QPO)
frequency rises with luminosity (Cui et al. 1999; Sobczak et al. 2000; 
McClintock et al. 2009).

\end{document}